\documentclass[aps,pra,twocolumn,showpacs,superscriptaddress]{revtex4}
\usepackage{graphicx}
\usepackage{amsmath}
\usepackage{amssymb}
\usepackage{color}
\newcommand{\HG}{\textup{HG}}
\newcommand{\LG}{\textup{LG}}

\graphicspath{{converted_graphics/}}

\begin{document}
\title{Decomposing spatial mode superpositions with a triangular optical cavity}

\author{G. H. dos Santos}
\affiliation{Departamento de F\'isica, Universidade Federal de Santa Catarina, Florian\'opolis, SC, 88040-900, Brazil}

\author{D. C. Salles}
\affiliation{Departamento de F\'isica, Universidade Federal de Santa Catarina, Florian\'opolis, SC, 88040-900, Brazil}

\author{M. G. Damaceno}
\affiliation{Departamento de F\'isica, Universidade Federal de Santa Catarina, Florian\'opolis, SC, 88040-900, Brazil}

\author{B. T. Menezes}
\affiliation{Departamento de F\'isica, Universidade Federal de Santa Catarina, Florian\'opolis, SC, 88040-900, Brazil}
\affiliation{Centro Brasileiro de Pesquisas F\'isicas, Rio de Janeiro, RJ, 22290-180, Brazil}

\author{C. Corso}
\affiliation{Departamento de F\'isica, Universidade Federal de Santa Catarina, Florian\'opolis, SC, 88040-900, Brazil}
\affiliation{Institute of Physical Chemistry, Polish Academy of Sciences, Kasprzaka 44/52, 01-224 Warsaw, Poland}

\author{M. Martinelli}
\affiliation{Instituto de Física, Universidade de São Paulo, São Paulo, SP, 05508-090, Brazil}

\author{P. H. Souto Ribeiro} 
\affiliation{Departamento de F\'isica, Universidade Federal de Santa Catarina, Florian\'opolis, SC, 88040-900, Brazil}

\author{R. Medeiros de Ara\'ujo}
\affiliation{Departamento de F\'isica, Universidade Federal de Santa Catarina, Florian\'opolis, SC, 88040-900, Brazil}

\date{\today}
\begin{abstract}
We demonstrate that a triangular cavity can operate as an efficient discriminator for Hermite-Gaussian optical modes. The process is applied to distinct fields produced by a Spatial Light Modulator, decomposing the generated paraxial field onto a basis of modes defined by the resonant 
optical cavity. We also present a proposal for the realization of a mode sorter based on a sequence of these cavities, without the need for optical isolation. We discuss its possible applications in sensitive measurements and distribution of information in optical systems. The high purity of the process should enable the use at the level of quantum information sharing.
\end{abstract}
\maketitle
%
\section{Introduction}

The degrees of freedom (DoF) of light, such as polarization, path, spatial mode and time/frequency mode, constitute important physical bases to encode information. Probably the most ubiquitous example is the internet traffic around the globe, which is mainly carried by light traveling in undersea optical-fiber cables. Today, pulsed lasers digitally encode the vast majority of the content humanity consumes every day \cite{Willner2019}. This is an example of information encoded on time-bin modes of light via amplitude modulation. 

At the single-photon level, DoF-encoded information becomes quantum, in the sense that the rules of quantum mechanics -- such as quantum superposition and interference -- become unavoidable when the information is transmitted and processed, or even interpreted. The description and interpretation of quantum information are, therefore, intrinsically attached to the quantum description of the system carrying information. For instance, the polarization DoF of a single-photon constitutes a basis for encoding a quantum bit, or q-bit \cite{Vernaz-Gris2018}. DoFs with higher dimensionality open the possibility to encode q-trits (three dimensions) or, more generally, q-dits ($d$ dimensions)\cite{Zang19}. The latter can be implemented using path \cite{Rossi09,Almeida2007}, spatial mode \cite{Walborn10,Neves2005} or time/frequency \cite{Brendel99,Marcikic02,Martin13,Kuzucu2005}, for instance. Orbital angular momentum of light (OAM) \cite{Allen92,Friese95,Yao11} is an infinite-dimension spatial DoF often referred to as a viable candidate for implementing q-dits, given the relatively well-developed classical toolbox for generating, coupling and measuring modes carrying OAM \cite{Giordani2019}. Increasing dimensionality in quantum communication and processing brings several possibilities of improvement. For instance, the use of high dimension alphabets in quantum cryptography increases security and the information transfer rate\cite{Walborn2006,Mufu13}. 

In this work, we demonstrate the use of a triangular cavity as a spatial-mode discriminator \cite{Berkhout10,Forbes16}. We use a spatial light modulator to prepare superpositions of Hermite-Gaussian modes and send them through the cavity. Tuning the cavity resonances allows the separation of the spatial modes. We also present a proposal of an optical scheme that realizes a mode sorter based on a cascade of cavities, 
with advantage of lower losses than those expected from cascaded linear cavities \cite{Wei2020}. The triangular geometry proposed here is particularly suited to the sorting scheme due to the easy access to the reflected beam.

The article if organized as follows: in sec. \ref{sec:modes} we review the Hermite and Laguerre-Gauss decomposition of paraxial beams, and their generation by optical masks is discussed in sec. \ref{sec:masks}. The properties of the proposed optical cavities are discussed in sec. \ref{sec:gouy}. The details of the setup are presented in Sec. \ref{sec:exp}, and the analysis of the results, in sec. \ref{sec:results}, are followed by the conclusions of the article.


\section{Hermite- and Laguerre-Gauss modes}
\label{sec:modes}

Among all possible bases of paraxial modes suitable for describing beams of light, the Hermite-Gauss and the Laguerre-Gauss bases are probably the most well-known.

The field complex envelope of a Hermite-Gaussian mode propagating along the $z$-axis with a waist at $z=0$ is described by \cite{Kogelnik1966}:
\begin{align}
&u^\textup{HG}_{mn}(x,y,z)\propto \frac{w_0}{w(z)}
H_m\left(\frac{\sqrt{2}x}{w(z)}\right)
H_n\left(\frac{\sqrt{2}y}{w(z)}\right)\nonumber\\
&\times\exp\left(-\frac{x^2+y^2}{w^2(z)}\right)
\exp\left(i\frac{k(x^2+y^2)}{2R(z)}-i\psi_{mn}(z)\right),
\label{eq:hermite}
\end{align}
where $w(z)$ is half of the beam diameter at $1/e^2$ of the peak intensity for a given $z$ ($w_0\equiv w(0)$ is the radius at the beam waist), $R(z)$ is the radius of curvature of the wavefront, $H_m$ are the Hermite polynomials and $\psi_{mn}(z)$ is the Gouy phase, which will be treated in the Section \ref{sec:gouy}.

The functions $u_{mn}^{HG}(x,y,z)$ form a basis of the space of square-integrable functions $L^2(\mathbb{R}^2)$. Physically, this means that any paraxial field $E(x,y,z)=u(x,y,z)\exp\{i(kz-\omega t)\}$ with finite energy and waist at $z=0$ can have its complex envelope decomposed as
\begin{equation}
u(x,y,z)=\sum_{m,n=0}^\infty c_{mn}u_{mn}^{HG}(x,y,z).
\end{equation}

The above reasoning also applies to the Laguerre-Gaussian modes, whose field complex envelopes write:
\begin{align}
&u^\textup{LG}_{p\ell}(\rho,\phi,z)\propto \frac{w_0}{w(z)}
\left(\frac{\sqrt{2}\rho}{w(z)}\right)^{|\ell|}
L_p^{|\ell|}\left(\frac{2\rho^2}{w^2(z)}\right)\nonumber \\
&\times\exp\left(-\frac{\rho^2}{w^2(z)}\right)
\exp\left(-i\frac{k\rho^2}{2R(z)}-i\ell\phi+i\psi_{p\ell}(z)\right),
\label{eq:laguerre}
\end{align}
where $p=0,1,2,...$ is known as the radial number and $\ell=0,\pm1,\pm2,...$ the azimuthal number, due to the azimuthal dependence of the phase term $i\ell\phi$, which provides the LG beam with a $\ell$-helical wavefront. $L_{p}^{|\ell|}$ are the associated Laguerre polynomials. It is now well known that in a Laguerre-Gauss beam of optical frequency $\omega/2\pi$, the ratio between OAM and intensity is equal to $\ell/\omega$ \cite{Beijersbergen1993}. This is why it is common to associate the quantity $\ell\hbar$ to the average amount of OAM per photon in a coherent beam.

LG modes may also be written in terms of HG modes (and vice-versa). The order of a mode is defined as $N=m+n$ for the Hermite-Gauss family and $N=|\ell|+2p$ for the Laguerre-Gauss families. In fact, the decomposition of a mode belonging to the $N$-th order of a given family decomposes exclusively onto modes of the same order $N$ of the other family.  In order to write the HG-decomposition of LG modes, it is convenient to change the usual $(p,\ell)$ indices by $(m,n)$ such that $p=\min(m,n)$ and $\ell=m-n$. Then \cite{Beijersbergen1993}:
\begin{equation}
u_{mn}^{LG}= \sum_{k=0}^N\,\frac{i^k}{\sqrt{2^N}}\, b_N(n,k)\,u_{N-k,k}^{HG}\ ,
\label{eq:LG-HG}
\end{equation}
where
\begin{equation}
b_N(n,k)=\!\sqrt{\frac{(N-k)!}{(N-n)!n!k!}}
\,\frac{d^k}{dt^k}[(1-t)^{N-n}(1+t)^n]_{t=0}.
\label{eq:b-coeffs}
\end{equation}
Inverting equation \eqref{eq:LG-HG} yields:
\begin{equation}
u_{mn}^{HG}=\sum_{k=0}^N\,\frac{(-i)^n}{\sqrt{2^N}}\, b_N(k,n)\,u_{N-k,k}^{LG}\ .
\label{eq:HG-LG}
\end{equation}
Note that one can take the factor $(-i)^n$ out of the summation as a global phase factor. This means that all LG components of a HG mode are all in phase (up to a phase difference of $\pi$ for some modes depending on the sign of $b_N(n,k)$).

\section{Holographic masks}
\label{sec:masks}

A mask is a structure that modulates the phase and/or the amplitude of an incident laser beam. The simplest OAM mask one can imagine multiplies the incident beam by a phase factor $\exp(i\ell\phi)$ while leaving the intensity profile unchanged. Consequently, although the transmitted beam has a perfectly helical phase structure, the intensity profile does not match that of a pure LG mode. However, a value of purity can be attributed to the beam by decomposing it in the LG basis. The Spiral Phase Plate (SPP) \cite{Beijersbergen1994} is a good example of device working as such simple OAM mask.

A slightly older method for generating beams with helical wavefronts uses holography. Holographic OAM masks are designed from interference patterns between a reference beam and a target beam, so that if a beam with a zero-order mode impinges on the mask, the resulting field acquires topological features matching the target \cite{Soskin1990}.  
Phase masks can also be implemented via a Spatial Light Modulator (SLM), a device with a liquid crystal display whose pixels change its refraction index according to the tension applied. SLMs are nowadays commonly adopted to create programmable, versatile an adjustable OAM masks.

When using SLMs, it turns out that the mode purity is improved with holographic masks, as compared to a simple azimuthal phase mask. This happens because the modulation efficiency of the device is limited, so that some fraction of the field amplitude is never modulated, decreasing purity on SLM-implemented SPPs \cite{Lemos14}. In holographic masks, however, the non-modulated part of the beam is discarded along the zeroth order of diffraction, optimizing purity on the first order. 
Each diffraction order may have its own mode-purity properties.

\begin{figure}
\centering
   \includegraphics[width=\columnwidth]{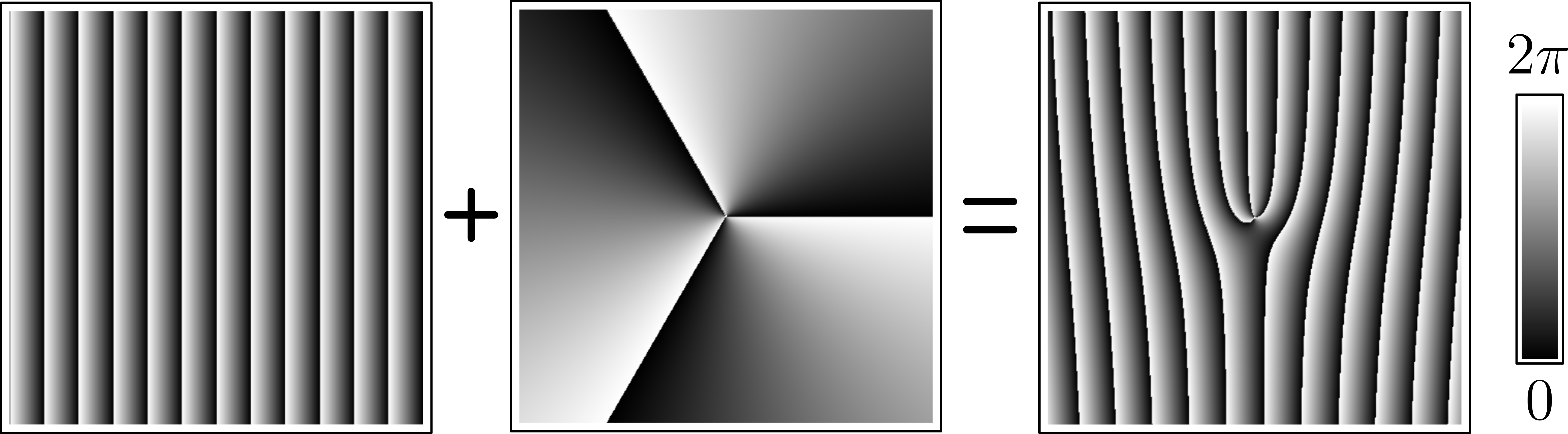}
   \caption{An OAM holographic mask ($\ell=3$, for example) can be understood as the sum (mod $2\pi$) of a blazed grating plus an azimuthal phase profile.}
   \label{fig:forks}
\end{figure}

A holographic mask designed from the interference of a zeroth-order Gaussian beam with a helical-wavefront beam is called a forked hologram, for its resemblance to a pitchfork. These holograms can also be described as ruled diffraction gratings -- for their seemingly horizontal periodicity -- added with an azimuthal dependence of $\ell$, as illustrated in Figure \ref{fig:forks}.

In order to design a forked hologram mask with similar properties to a blazed grating, one models the phase modulation with the sawtooth function of amplitude $2\pi$ over a period $\Lambda$. The transmittance is then
\begin{equation}
\label{BlazedForkedHologram}
    T(x,y)=e^{i(kx-\ell\phi(x,y))}.
\end{equation}
where $\phi(x,y)$ is the azimuthal angle, $k=2\pi/\Lambda$ and $\Lambda$ can be chosen according to the wavelength $\lambda$ and the desired angle of diffraction.
This blazed forked mask is ideal for SLMs because it not only avoids the limited modulation capacity of the device, but it also maximizes the power coupled to the target mode.

So far, we have considered phase-only modulation. Mode purity achieved with such holographic masks can still be improved by adding some Gaussian amplitude modulation, so that the amplitude profile of the output beam is even closer to that of a pure LG mode. In fact, as far as only OAM (index $\ell$) is concerned, non-Gaussian masks are sufficient, but they actually generate a particular superposition of modes with same $\ell$ and different $p$ indices. As our need for information processing capacity increases, being able to produce arbitrary LG modes, varying $\ell$ and $p$ independently, becomes an important tool.

\section{Optical cavity}
\label{sec:gouy} 

An optical cavity is an interferometer composed of a set of mirrors defining a certain round-trip length. When the optical phase accumulated over a round-trip is a multiple of $2\pi$, constructive interference occurs inside the cavity, maximizing optical intensity at its outputs: light and cavity are said to be in resonance. Far from resonance (destructive interference) the light is reflected upon the input mirror. 

By scanning the optical frequency or the cavity length, it is possible to observe intensity peaks revealing the resonances. These peaks are separated, in frequency, by the inverse of round-trip time of flight or, in length, by the wavelength $\lambda$.

\textbf{Linear cavity.} In an ideal cavity with two plane mirrors facing each other at a distance $L$, the resonances with a plane wave of wavelength $\lambda$ would occur for cavity round-trip lengths $2L=p\lambda$, where $p$ is a positive integer. 

This arrangement, however, is not geometrically stable, causing a realistic laser beam to diverge eventually. A realistic geometrically stable cavity, must have at least one optical element with finite focal length in order to encompass Gaussian modes. Generally, concave mirrors do this job, compensating for the round-trip beam divergence.

As it propagates, a Gaussian beam accumulates a geometrical phase shift called Gouy phase, first observed with thermal light propagating through a focal point in 1890 by French physicist Louis Georges Gouy \cite{Gouy1890}. The Gouy phase $\psi_N(z)$ of a $N^\textup{th}$-order laser mode writes \cite{Saleh1991}:
\begin{equation}
\psi_N(z)=(N+1)\tan^{-1}(z/z_R)\ ,
\label{eq:gouy-phase}
\end{equation}
where $z$ is measured with respect to the location of the beam waist and $z_R=\pi w_0^2/\lambda$ is the Rayleigh length.

This phase shift causes a displacement $\varepsilon_N$ on the position of the resonance peaks as compared to the case of plane waves, meaning that the $N^\textup{th}$-order peaks actually occur at cavity round-trip lengths $2L=p\lambda+\varepsilon_N$.

For a cavity composed of a plane mirror and a concave mirror at a distance $L$ from each other, computing the accumulated Gouy phase over a cavity round-trip yields
\begin{equation}
    \varepsilon_N=
    (N+1)\ \frac{\tan^{-1}(L/z_R)}{\pi}\ \lambda\ .
    \label{eq:deltaN}
\end{equation}

\begin{figure}[h!]
\centering
   \includegraphics[width=1\columnwidth]{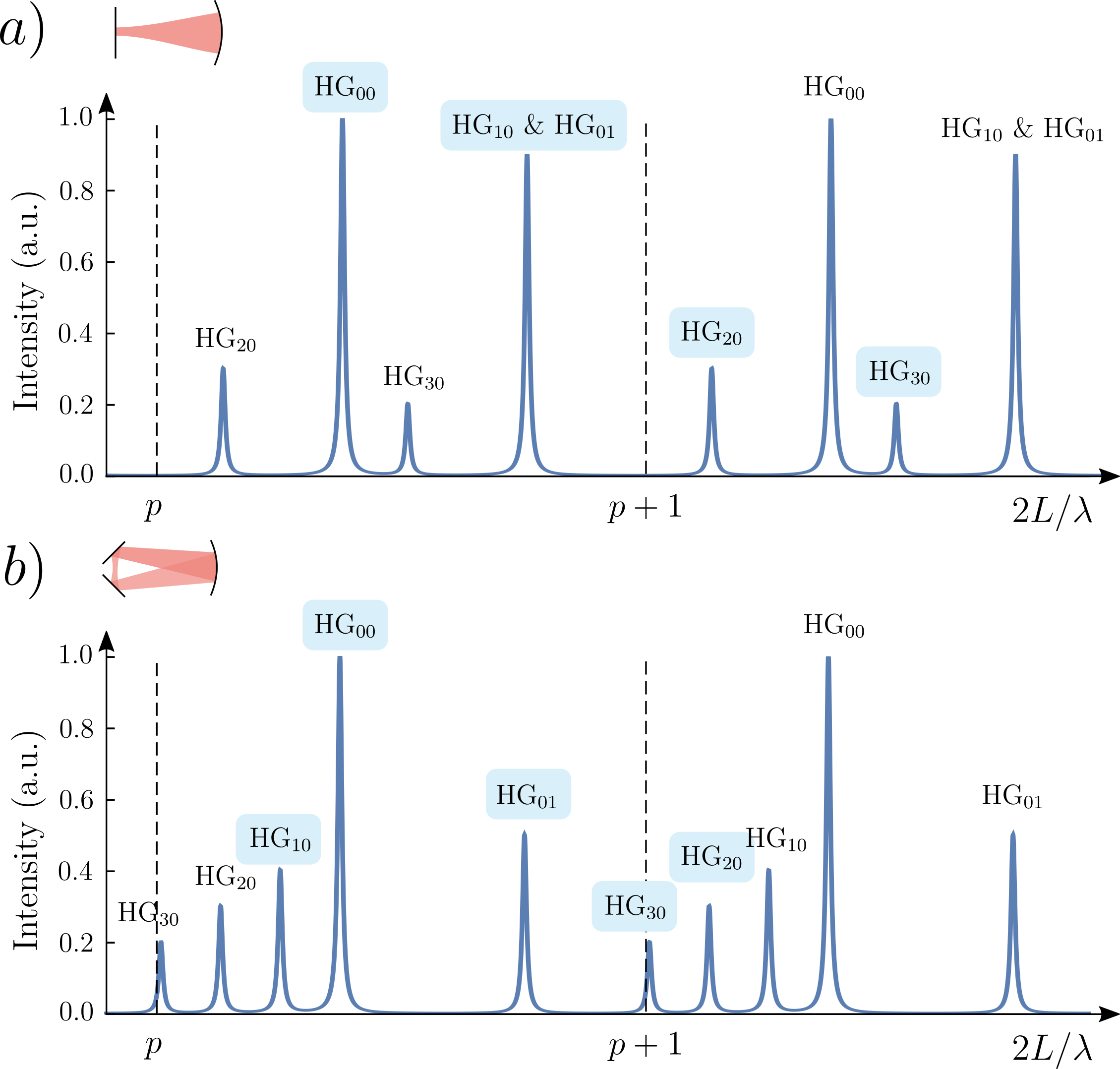} 
   \caption{Resonance peaks of a superposition of the modes $\HG_{00}$, $\HG_{10}$, $\HG_{01}$, $\HG_{20}$ and $\HG_{30}$ (proportion 10 : 5 : 4 : 3 : 2) as one microscopically scans the total length $2L$ of (a) a linear cavity and (b) a triangular cavity. The simulation considers the same total length ($2L= 34.4\,$cm) and the same mirror curvature for both cavities ($R_c= 20\,$cm). These numbers correspond to a regular spacing $\varepsilon_{N+1}-\varepsilon_{N}\simeq0.38\lambda$ in a linear cavity and were chosen as realistic parameters matching the geometrical features of the triangular cavity used in the experiment.}
   \label{fig:peaks-theory}
\end{figure}

Note that the peak displacement $\varepsilon_N$ increases linearly with $N$. As a result, there is a regular spacing between the resonant peaks corresponding to the different orders $N$, as illustrated in Figure \ref{fig:peaks-theory}a.

\textbf{Triangular cavity.} In cavities with an even number of mirrors, the total phase accumulated in a cavity round trip is the same for all modes belonging to a given order. However, this may not be the case for cavities with an odd number of mirrors.

A reflection on a standing mirror flips the orientation of the $x$ axis, which implies a sign change on both horizontal coordinate and horizontal unit vector: $x\rightarrow -x$ and $\hat{x}\rightarrow -\hat{x}$ \cite{Sasada2003}. Thus, for example, the electric field of a $\HG_{10}$ mode (antisymmetric on the $x$ variable) with vertical polarization acquires a minus sign (i.e. an extra $\pi$ phase) for each reflection. After $n$ reflections, the extra acquired phase is $n\pi$. Therefore, for a cavity with an odd number of mirrors, the extra phase per round-trip due exclusively to reflections  equals $\pi$ (mod $2\pi$) for such mode (and 0 for $x$-symmetric modes).

Because of the exposed above, a triangular cavity is able to discriminate between the modes $\HG_{10}$ and $\HG_{01}$. The distance between their peaks is half the distance between two $\HG_{00}$ consecutive peaks. That corresponds to half a free spectral range (FSR) in frequency, or to $\lambda/2$ in total cavity length.

Figure \ref{fig:peaks-theory}b demonstrates how the difference between a linear and a triangular cavity manifests in terms of resonance peaks.

\section{Experimental Setup}
\label{sec:exp}

\begin{figure}
\centering
   \includegraphics[width=\columnwidth]{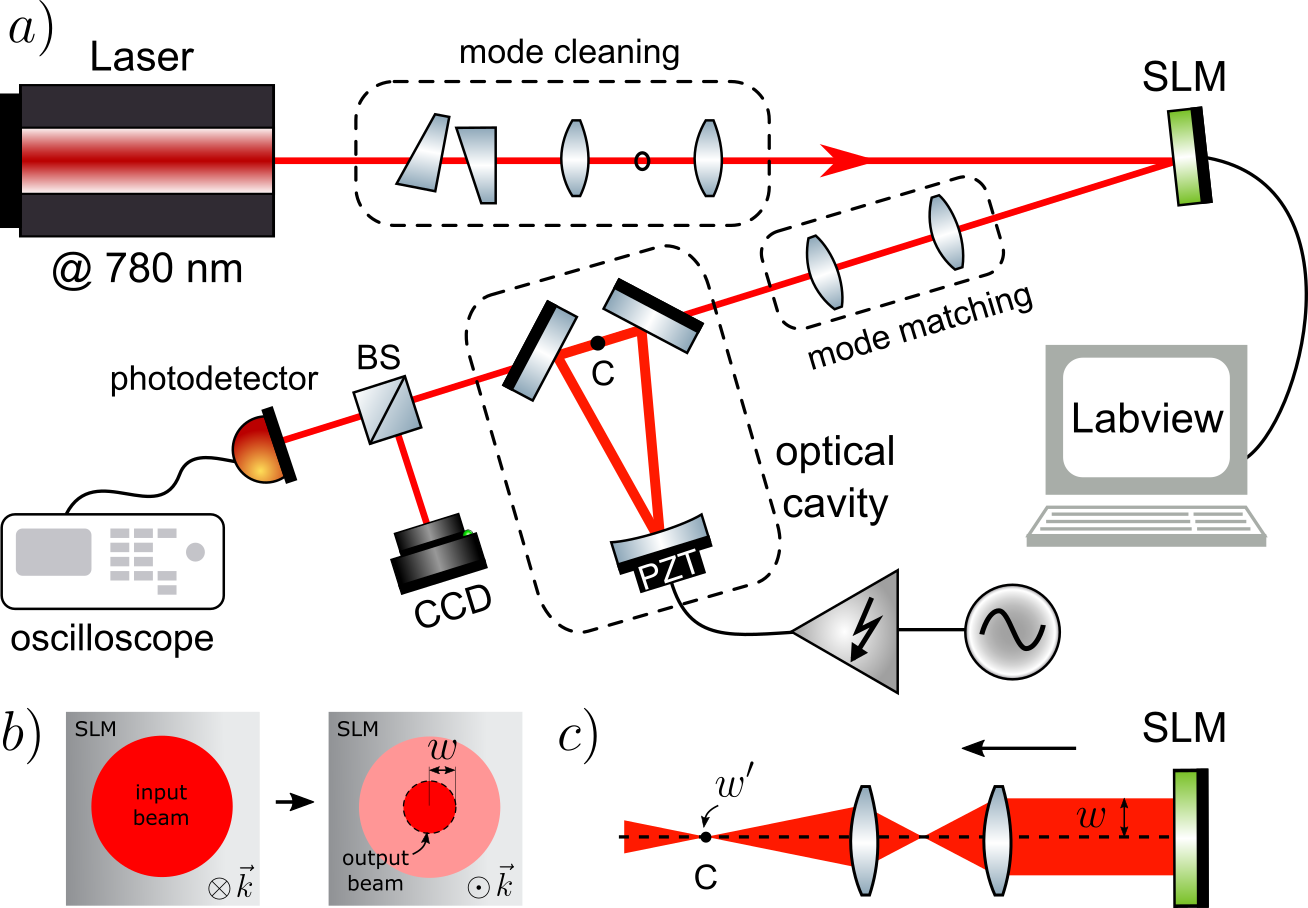}
   \caption{a) Experimental setup composed by a laser, a Spatial Light Modulator (SLM) controlled via Labview and an optical cavity whose length is swept by a piezoelectric transducer (PZT); b) SLM Gaussian mask to control the output waist; c) Mode matching scheme.}
   \label{fig:setup}
\end{figure}

The main elements of the setup are a diode laser as the source, a Spatial Light Modulator (SLM) and a triangular optical cavity, as shown in Figure \ref{fig:setup}. Prior to being sent to the SLM, the laser beam is mode cleaned with a pinhole of the appropriate size to ensure a TEM$_{00}$ Gaussian mode. The SLM then creates a superposition of optical modes to be analyzed by the cavity. As the cavity length is scanned over time with a frequency of $\sim 10$ Hz, a photodetector connected to an oscilloscope collects the transmitted light, capturing the resonances of the different Hermite-Gauss modes that compose the beam. The Hermite-Gauss basis used to describe the incoming beam is the family of resonant modes of the optical cavity, of waist $w'_0$. 

The cavity is composed by two identical partially transmissive plane mirrors and one concave mirror (radius of curvature $R_c=20\,$cm). The measured finesse is $\sim 100$ for $s$-polarized light. From $R_c$ and the cavity round-trip length $2L$, it is easy to calculate the Rayleigh length as $z_R=\sqrt{(R_c-L)L}$. In our experiment, since $L=17.2\,$cm and $\lambda=780\,$nm, we obtain $z_R=6.9\,$cm and a resonant beam waist $w'_0=131\,\mu$m located in between the plane mirrors, at point C. Astigmatic effects due to the non-zero angle of incidence $\theta$ on the concave mirror are relatively small and may be neglected in first approximation. Indeed, in our case, $\theta=5^\circ$, leading to calculated beam waists of 130 and 132 $\mu$m in the tangential (horizontal) and sagittal (vertical) directions, respectively. This astigmatism corresponds to a peak displacement of less than 0.1\% of the FSR.

Let us call $w$ the waist of the collimated beam leaving the SLM (Fig. \ref{fig:setup}b) and $w'$ the waist of the beam after the mode-matching lenses, at point C. Beam alignment and mode-matching are carefully adjusted, until $w'=w'_0$. For this procedure, no particular superposition of modes is created: the SLM is set as a simple blazed grating with a Gaussian amplitude modulation, sending a TEM$_{00}$ mode to the cavity on the first order of diffraction.

The mode matching lenses perform a linear scaling transformation around the optical axes, reducing the size of beam waist by a factor 7 from the SLM plane to the plane containing point C, inside the cavity. By convention, we denote the waist sizes at the cavity plan with a prime ($w'$; $w'_0$) and those at the SLM without a prime ($w$; $w_0$).

Once alignment and mode matching are optimized, one can then switch masks and create a mode superposition with the SLM. In order to appropriately read the resonance peaks, we use the TEM$_{00}$ peak as a reference and calculate the predicted distance to higher-order peaks using the geometrical features of the cavity (parity of the number of mirrors, length $L$ and mirror radius of curvature $R_c$).

\section{Experiments and Results}
\label{sec:results}

In order to analyze the performance of the cavity as a mode discriminator, we have used the following strategy. We prepare spatial mode superpositions and analyze them with the cavity. The way the superpositions are prepared is based in the use of simple phase masks in the SLM that couples the input zero-order $\HG_{00}$ mode to higher-order $\HG_{mn}$ modes.

\begin{figure}
\centering
   \includegraphics[width=\columnwidth]{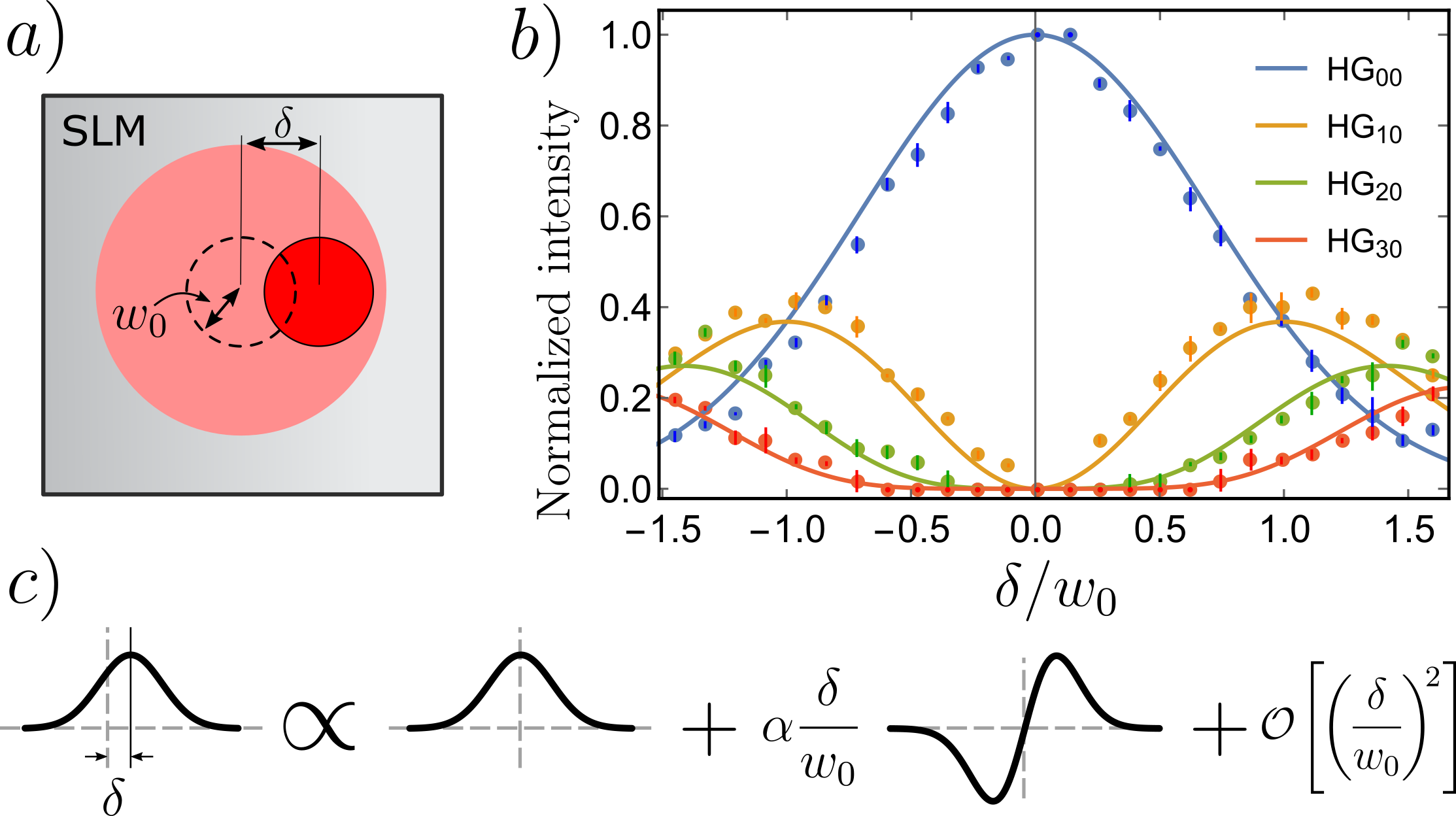}
   \caption{Misaligned zero-order Gaussian mode. a) Scheme of the SLM mask; b) Detected peak-intensity for each mode as a function of $\delta/w_0$, where $\delta$ is the horizontal mask misalignment relative to the beam axis. Dots are experimental data and solid lines are theoretical predictions. Blue: $\HG_{00}$ mode; orange: $\HG_{10}$ mode; green: $\HG_{20}$ mode; red: $\HG_{30}$ mode. c) Decomposition of a displaced Gaussian function on the basis of centered Hermite-Gaussian functions. The Gaussian with a small displacement $\delta$ as compared to the waist $w_0$ is essentially the superposition of the centered Gaussian and the first-order HG function.}
   \label{fig:result-align}
\end{figure}

Figure \ref{fig:result-align} shows the results of an experiment using the SLM to simulate a misalignment between the beam and the cavity. The beam of waist $w_0$ leaving the SLM arrives to the cavity with the beam waist $w'_0$ of its resonant modes. One can however control the lateral deviation $\delta'$ of the beam propagation axis with respect to the cavity axis. When $\delta'$ is zero, the beam is perfectly aligned and mode-matched to the cavity. As the $\delta'$ increases to positive values, the beam is horizontally dislocated, say, to the right and, for negative values of $\delta'$, the beam is dislocated to the left. The deviation $\delta'$ is implemented with a Gaussian mask of fixed radius whose center is scanned horizontally, deviating $\delta$ from the origin. The deviation $\delta'$ and $\delta$ are also subject to the linear scaling of the mode-matching, so that $\delta/\delta'=w_0/w'_0\approx 7$. The Gaussian mask is applied to an initial beam with very large waist impinging the SLM ($\sim2\times w_0$).


The results synthesized in Figure \ref{fig:result-align}b show that, even if the deviated beam remains Gaussian, higher-order modes are detected by the cavity. As the deviation $\delta$ increases relative to $w_0$, the contribution of the fundamental mode decreases. For small deviations, the first mode to take its place is the first-order mode $\HG_{10}$. In sequence, the modes $\HG_{20}$ and $\HG_{30}$ also appear and the contribution of $\HG_{10}$ is maximal for $\delta=w_0$. There is a good quantitative agreement between theory and experiment.

While the data points are obtained by measuring the relative heights of the cavity resonant peaks, each theoretical curve of Figure \ref{fig:result-align}b is the square of the inner product between the input mode and a resonant HG mode. Figure \ref{fig:result-align}c, for instance, shows that the inner product between a displaced Gaussian function and the first Hermite-Gaussian function is proportional to the displacement $\delta$ for small displacements ($\delta\ll w_0$). Thus, the square of the inner product under this condition is proportional to $\delta^2$. Indeed, the yellow curve on Figure \ref{fig:result-align}b can be manifestly well approximated by a parabola around $\delta=0$.  

The following experiment was designed to investigate what happens when one varies the waist $w$ of the beam sent to the cavity, whose resonant modes are compatible with an SLM beam waist $w_0$. This experiment implements a mode mismatch, while keeping the beam and the cavity perfectly aligned. Gaussian masks of adjustable radii are implemented on the SLM, causing $w$ (and thus $w'$, at point C) to change. Due to mode-matching linear scaling, $w'$ and $w$ vary proportionally.

\begin{figure}
\centering
   \includegraphics[width=\columnwidth]{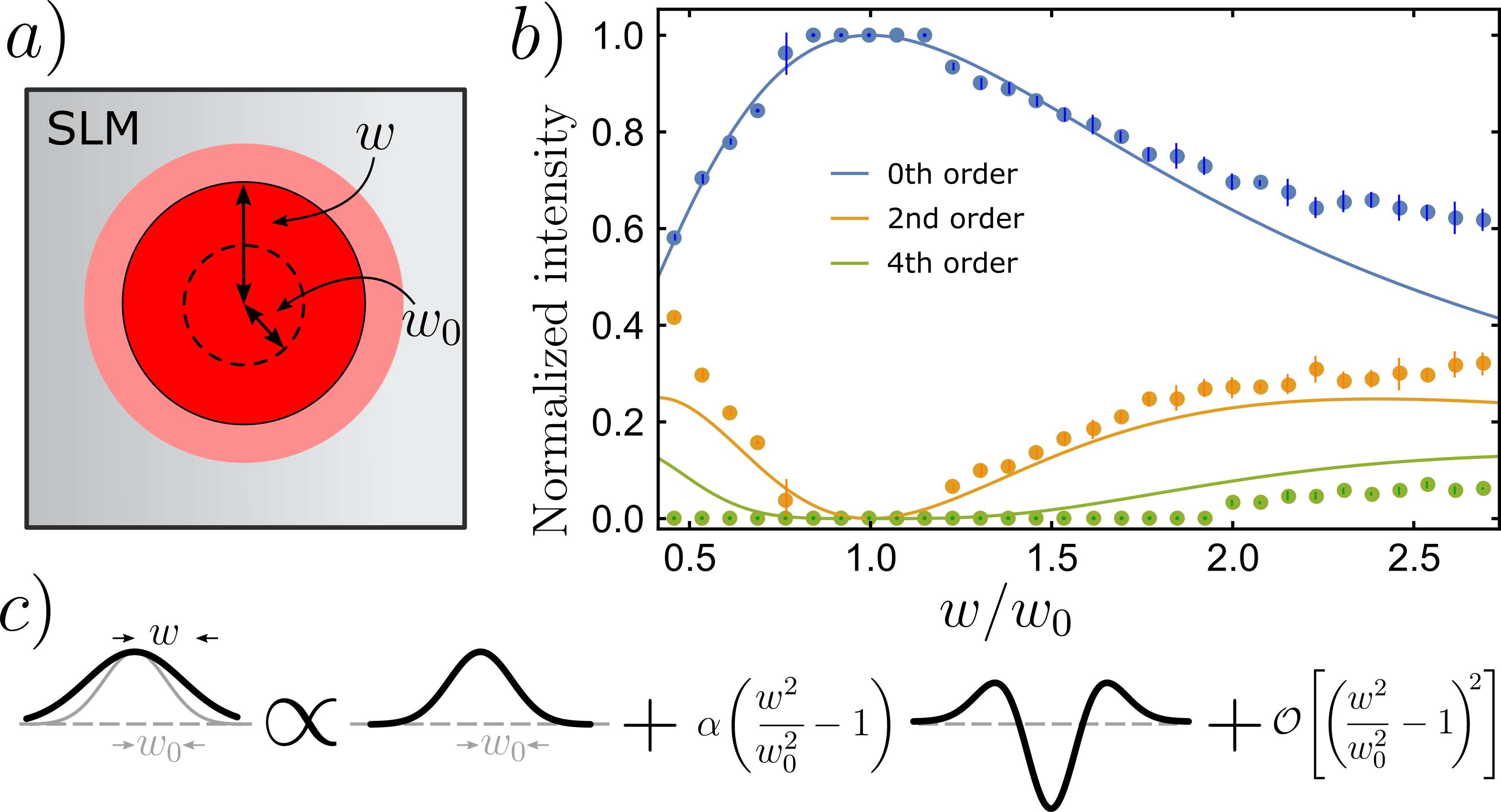}
   \caption{Mismatched Gaussian mode. a) Scheme of the SLM mask; b) Detected peak-intensity for each mode order as a function of $w/w_0$, where $w$ is the waist of the beam leaving the SLM and $w_0$ is the ideal SLM beam waist for cavity mode-matching. Dots are experimental data and solid lines are theoretical predictions. Blue: $\HG_{00}$ mode; orange: 2$^\textup{nd}$-order modes; green: 4$^\textup{th}$-order modes; c) Decomposition of a Gaussian function of width $w$ on a basis of Hermite-Gaussian functions of width $w_0$. When the ratio $w/w_0$ is sufficiently close to 1, the mismatched Gaussian is essentially the superposition of the zero-order Gaussian and the second-order HG function.}
   \label{fig:result-MM}
\end{figure}

Figure \ref{fig:result-MM} displays the results of this experiment. Differently from the previous one, the electric field of the input modes now show symmetry of rotation around the cavity axis. As a result of this symmetry, only even-order cavity modes participate in the decomposition. This is why the curves presented in Figure \ref{fig:result-MM}b refer to orders, not modes. In principle, a given order could be split into two resonant peaks, but it is not the case in this particular example. Thus, each curve in Figure \ref{fig:result-MM}b represents the height of a single peak. One can deduce from Figure \ref{fig:result-MM}c that the first three terms of the decomposition of a size-mismatched Gaussian beam are $\HG_{00}$, $\HG_{02}$ and $\HG_{20}$, all even functions on the variable $x$. The mode $\HG_{11}$, which is an odd function of $x$, is not involved in this particular decomposition, even if it is a 2$^\textup{nd}$-order mode. The experimental results corroborate this prediction, since no peak for $\HG_{11}$ is observed. The graphic of results, which are in good qualitative agreement with theory, shows that the second-order component appears whether the beam waist is larger or smaller than $w_0$.

Let us now study the performance of common masks utilized to produce Hermite- and  Laguerre-Gaussian modes of the first order, namely $\{\HG_{10},\,\HG_{01}\}$ and $\{\LG_{01},\,\LG_{0-1}\}$ modes. We also investigate the influence of the horizontal misalignment of a mask on the purity of the output mode.

\begin{figure}
\centering
   \includegraphics[width=\columnwidth]{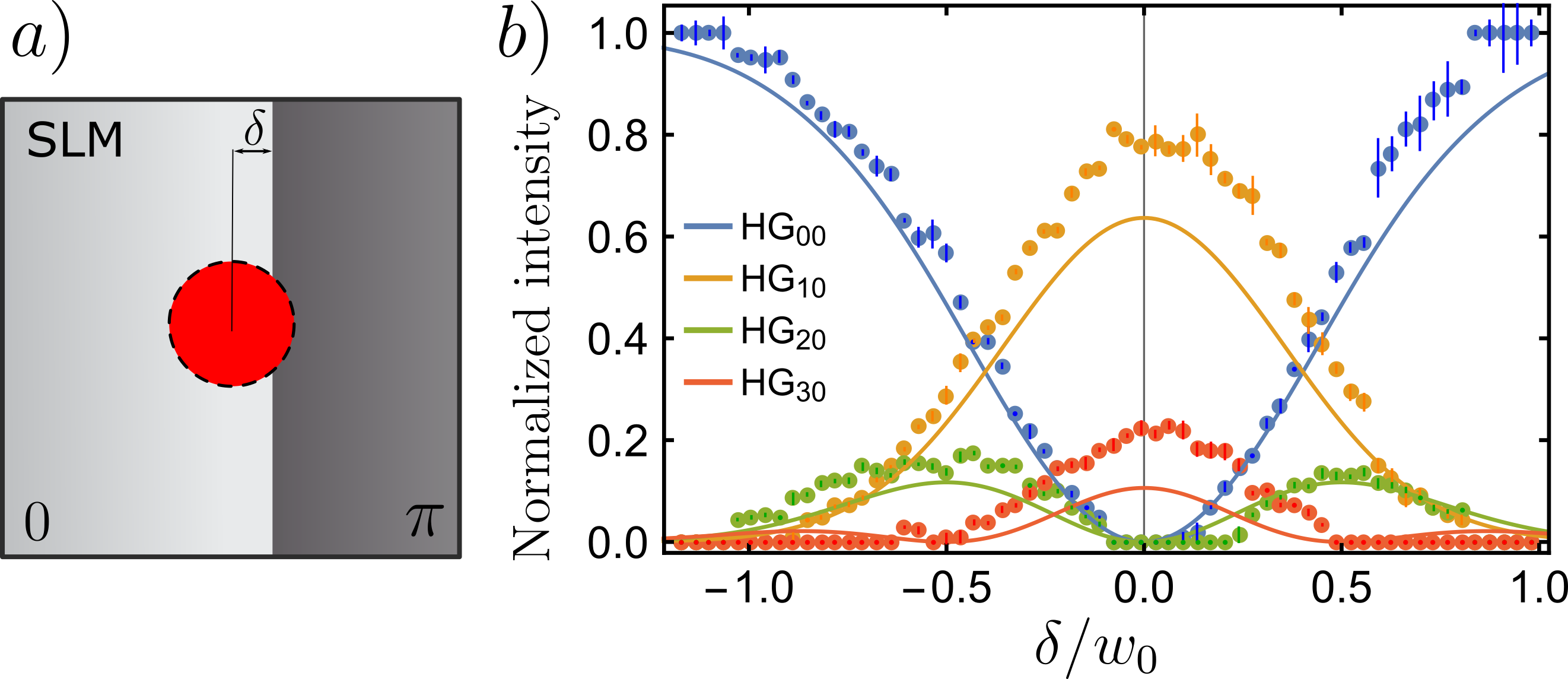}
   \caption{a) Misaligned mask for $\HG_{10}$; b) Detected peak-intensity for each mode as a function of $\delta/w_0$, where $\delta$ is the horizontal mask misalignment relative to the beam axis. Dots are experimental data and solid lines are theoretical predictions. Blue: $\HG_{00}$ mode; orange: $\HG_{10}$ mode; green: $\HG_{20}$ mode; red: $\HG_{30}$ mode.}
   \label{fig:result-HG10}
\end{figure}

We start with a Hermite-Gaussian mask, whose non-blazed component consists of two flat-phase regions, with phases 0 and $\pi$, respectively, separated by a vertical straight line, as represented in Figure \ref{fig:result-HG10}a. This mask mimics the phase profile of a $\HG_{10}$ mode. A zero-order Gaussian beam impinging onto such mask will display at the immediate output a first-order phase profile while keeping a zero-order Gaussian intensity profile. Although it leads to a mere approximation of an actual $\HG_{10}$ mode, this mask can be useful in applications that do not demand high mode purity. Here, we measure the mode purity not only at ideal operation conditions of alignment, but also the effect of a lateral displacement $\delta$ on the mask, so that there is an adjustable distance $\delta$ between the vertical line and the center of the impinging beam.

Theory predicts that the mode purity is nearly 65\% for $\delta$ = 0, as indicates the yellow solid line in Fig. \ref{fig:result-HG10}b). Despite the overall good qualitative agreement between experimental results and theoretical curves, we see that, for $\delta=0$, the measured purity is significantly higher than predicted: $\sim$80\%. This is due mainly to undetected higher-order modes that are buried under background intensity noise, overestimating the normalized intensity of the lower-order modes (see Sec. \ref{sec:discussion} for details).
Although not demonstrated experimentally, simple numeric calculations show that, by stretching the input beam waist on the $x$ direction by a factor 1.7, one can achieve a theoretical purity above 80\%.

\begin{figure}
\centering
   \includegraphics[width=\columnwidth]{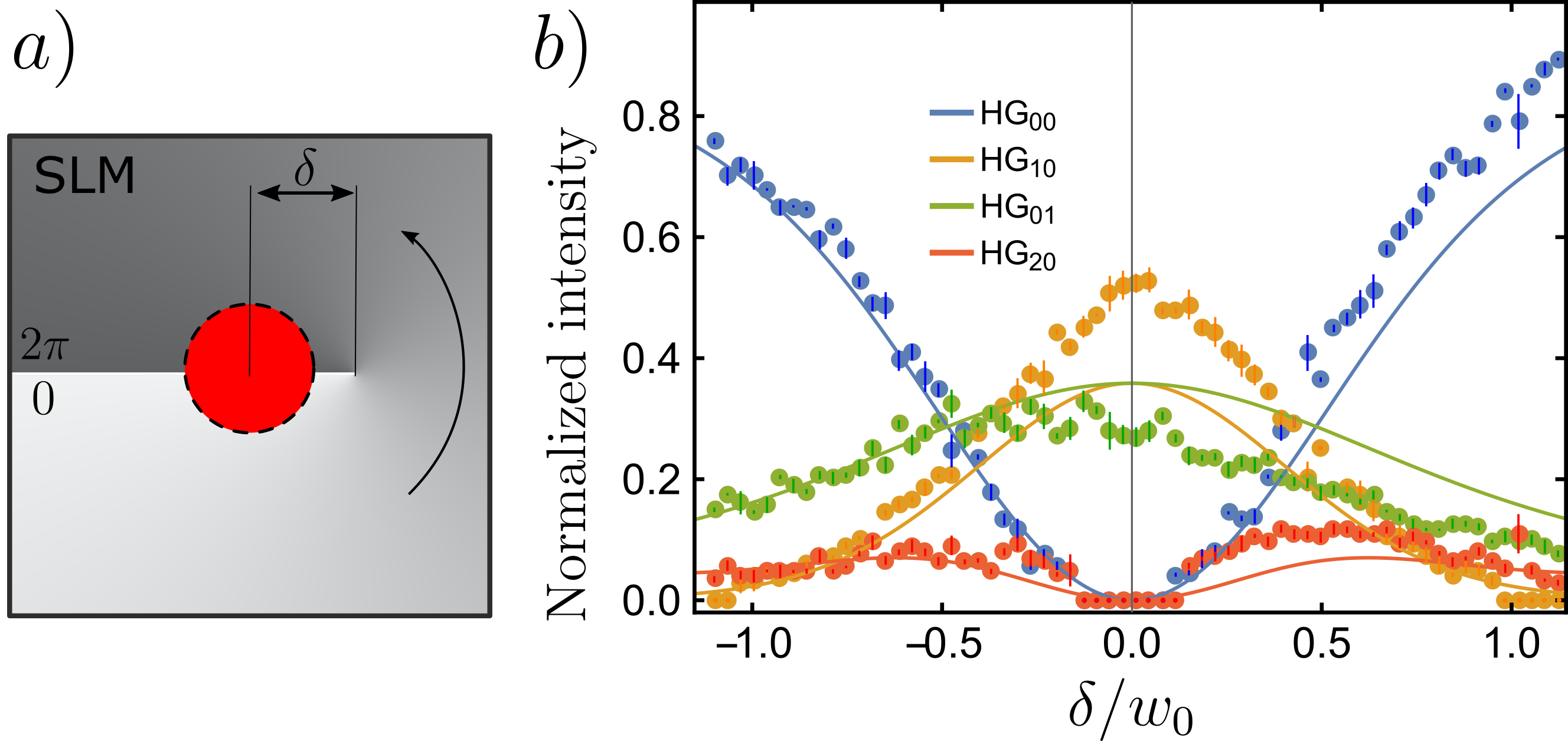}
   \caption{a) Misaligned mask for $\LG_{10}$; b) Detected peak-intensity for each mode as a function of $\delta/w_0$, where $\delta$ is the horizontal mask misalignment relative to the beam axis. Dots are experimental data and solid lines are theoretical predictions. Blue: $\HG_{00}$ mode; orange: $\HG_{10}$ mode; green: $\HG_{01}$ mode; red: $\HG_{20}$ mode; purple: $\HG_{02}$ mode; brown: $\HG_{30}$ mode.}
   \label{fig:result-LG10}
\end{figure}

Another common mask utilized for the generation of first-order modes displays an azimuthal phase component (in addition to the blazed phase component), as shown in Figure \ref{fig:result-LG10}a. This mask creates an approximation of an $\LG_{01}$ (or $\LG_{0-1}$, depending on whether phase increases clockwise or counterclockwise from 0 to $2\pi$) aligned to the first order of diffraction. Immediately after the reflection on the SLM, the beam phase profile corresponds to that of a $\LG_{01}$ mode, although the intensity profile is Gaussian. It is only upon propagation that the beam acquires a donut-like intensity profile.

Assessing the mode purity achievable experimentally with such a mask involves projecting the output light onto the mode $\LG_{01}$, which decomposes as $(\HG_{01}+i\HG_{10})/\sqrt{2}$ on the HG basis. The mode $\LG_{01}$, however, is not resonant to our triangular cavity, which displays distinct resonant peaks for vertical and horizontal first-order HG modes. Indeed, because of the extra $\pi$ phase acquired by $\HG_{10}$ on the third intracavity reflection every round trip, the peaks are as far as they can possibly be from each other: half a free spectral range. This is why we performed our analysis entirely on the HG basis.

For an input Gaussian beam perfectly aligned to the mask, it is reasonable to estimate the LG mode purity as the sum of the normalized peak intensities measured for the first-order HG modes, even though we are not able to measure the relative phase between these modes. In theory, the purity achieved with this mask with ideal alignment is $78.5$\%. It is possible to increase this number up to $93.1$\% if the input beam waist is expanded to $\sqrt{2} w_0$. Experimentally, the result is not conclusive, because of a visible mode unbalance between $\HG_{10}$ and $\HG_{01}$. The reasons for this unbalance are not clear, although a few hypothesis acting together could explain it: cavity astigmatism, input mode ellipticity and SLM phase miscalibration.

As we introduce a mask displacement $\delta/w_0$ relative to the SLM beam waist $w_0$, a stronger $\HG_{10}$ component is expected, and the experimental estimation of the LG-mode purity is no longer reliable, even if the results were in perfect quantitative agreement with theory. For significant displacements ($\delta\gtrsim 0.5w_0$), we see in Figure \ref{fig:result-LG10}b that the fundamental mode becomes dominant over the higher-order modes, since the center of the beam gets further from the mask phase singularity.

\begin{figure}
\centering
   \includegraphics[width=\columnwidth]{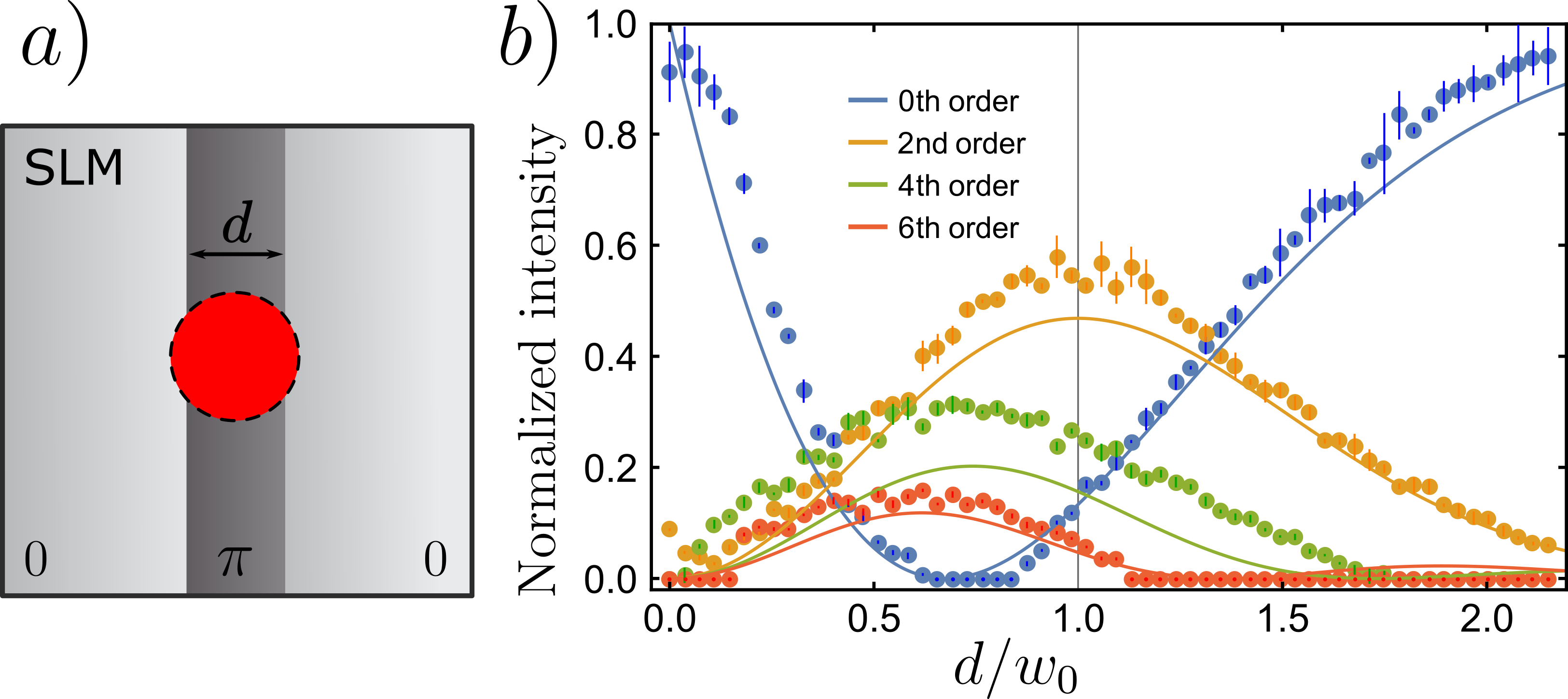}
   \caption{a) Phase mask for $\HG_{20}$; b) Detected peak-intensity as a function of $d/w_0$, where $d$ is the width of the central region of the $\HG_{20}$ mask (with applied phase $\pi$). Dots are experimental data and solid lines are theoretical predictions. Blue: $\HG_{00}$ mode; orange: $\HG_{20}$ mode; green: $\HG_{40}$ mode; red: $\HG_{60}$ mode.}
   \label{fig:result-HG20}
\end{figure}

The last mask tested in this experiment displays three separate regions with phases 0, $\pi$ and 0, respectively (Figure \ref{fig:result-HG20}a), simulating the phase profile of a $\HG_{20}$ mode. The central region has a tunable width $d$ on the SLM. By scanning $d$ from 0 up to $2w_0$, we compile the experimental data shown in Figure \ref{fig:result-HG20}b. We show that the $\HG_{20}$ component is maximum when $d=w_0$, reaching around 50 to 55\% of total intensity. On the limits $d\rightarrow 0$ and $d\rightarrow \infty$, the fundamental Gaussian mode tends to dominate, since the mask tends to display a flat phase over the entire beam profile
. 

\section{Discussion}
\label{sec:discussion}

The results presented in the previous section (graphics of figures \ref{fig:result-align} to \ref{fig:result-HG20}) are all in good qualitative agreement with theory, described by solid lines. It is worth noticing that the solid lines are ``pure'' theoretical predictions, with no parameters adjusted to experimental data.

A careful inspection of the results shows, however, some quantitative disagreement in several situations, which we look deeper into in this section.

First, on figure \ref{fig:result-MM}b, we notice that, for $w>2w_0$, the data points seem to stop evolving, while the distance to solid lines seems to increase. This can be easily explained by the fact that our input beam waist on the SLM is no larger than $2w_0$, setting an upper limit to the output radius.

In Figure \ref{fig:result-HG10}, the data points within the interval $|\delta/w_0|<0.5$ take values considerably above the theoretical curves. We believe this is due to undetected higher-order modes, whose resonant peaks are not clearly visible, since their sizes compare to the background intensity noise. In fact, the total intensity (sum over all modes detected) is normalized to 1 for each new setting of $\delta$. For $\delta=0$, theory predicts that $\HG_{10}$ and $\HG_{30}$ modes contribute together with 74.3\% of total intensity, while the next relevant modes -- $\HG_{50}$, $\HG_{70}$ and $\HG_{90}$ -- contribute with less than 5\%, 3\% and 2\% respectively. Because of the phase discontinuity, many higher-order modes contribute very little individually and convergence is considerably slow. A more refined data analysis would certainly help removing some of the background noise, but this is beyond the scope of the present work.

The same issues are present in Figure \ref{fig:result-HG20}, but their impact in the experimental data seems to be lower, probably because the phase discontinuities are rather far from the center of the beam.

Finally, on Figure \ref{fig:result-LG10}, one may notice an asymmetry of the experimental with respect to the $\delta=0$ axis. The quantitative agreement between theory and experiment is clearly better for $\delta<0$. The right side of the graphic ($\delta>0$) corresponds to situation where the beam impinges the mask over the 0-$2\pi$ line. Although 0 and $2\pi$ should be physically equivalent, our hypothesis is that there must have been a slight phase miscalibration on the SLM, causing this line to present a phase discontinuity that introduced contributions from higher-order modes.

Despite the overall good performance, an intrinsic limitation persists on the use of optical cavities to reveal the modal structure of paraxial beams. Although the triangular cavity is able to discriminate more HG modes than a linear cavity thanks to the extra reflection, some modes remain inseparable. To give an example, the order 2 have three modes, two of which are $x$-symmetric and resonate together, namely, $\HG_{20}$ and $\HG_{02}$. This degeneracy can be raised at the cost of analyzing the image at the output of the cavity, as we actually have done using the CCD camera shown in Figure \ref{fig:setup}.

\section{Mode sorter}
\label{sec:sorter}

In the previous sections, we have shown both theoretically and experimentally that a triangular optical cavity is a useful tool in the analysis of the mode structure of a paraxial beam. In this section, we show that it can also be used to build a mode sorter based on a cascade of cavities.

\begin{figure}[t]
\centering
   \includegraphics[width=\columnwidth]{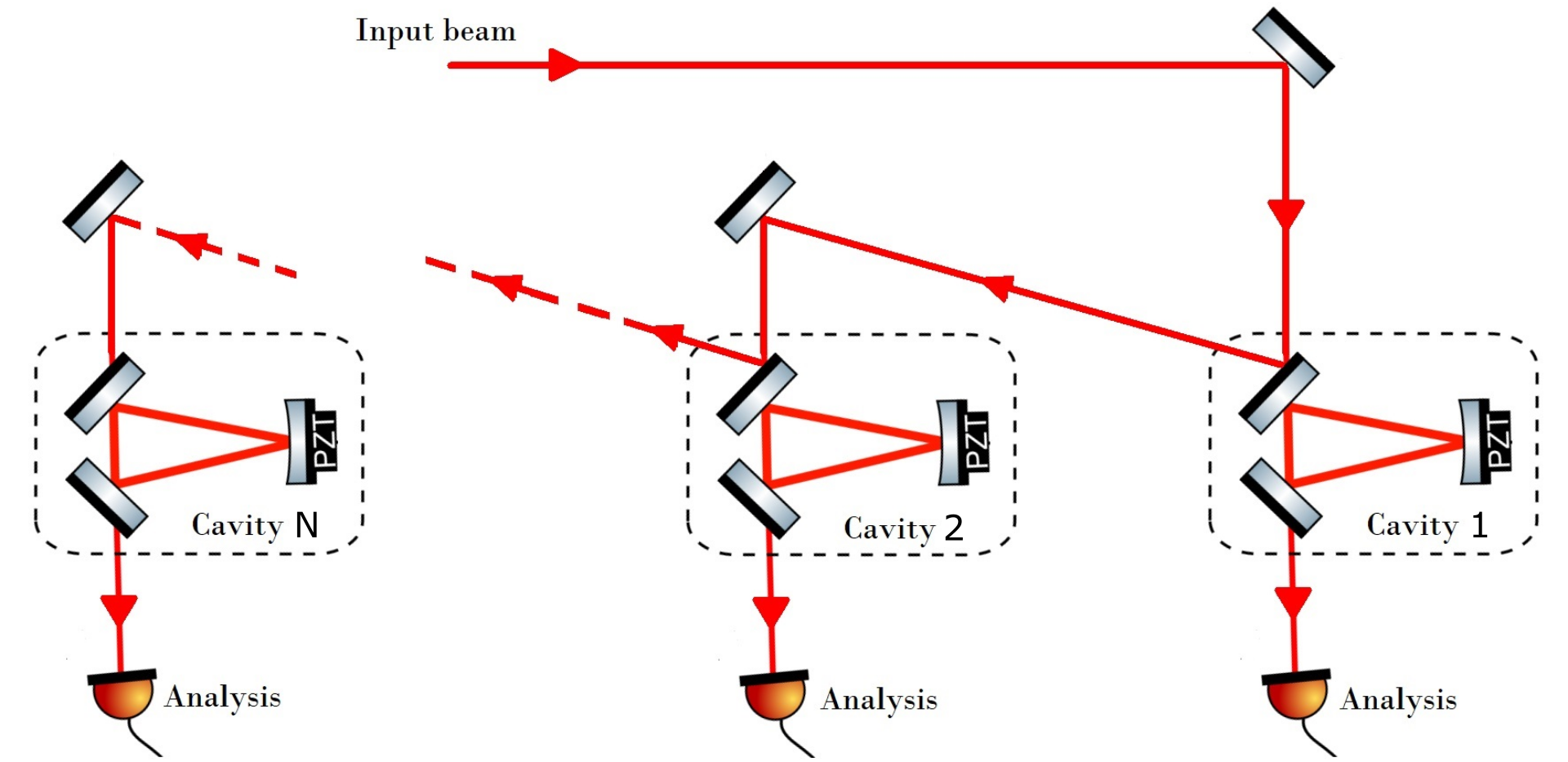}
   \caption{Experimental scheme proposed to realize a spatial mode sorter based on a sequence of triangular cavities. See details in the main text.}
   \label{fig:sorter}
\end{figure}



The proposed experimental scheme is shown in Fig. \ref{fig:sorter}. The input laser beam (some superposition of spatial modes) is incident in the input mirror of cavity 1, which is tuned and locked at resonance with some specific mode(s). The transmitted light is therefore filtered out in the selected mode(s). The reflected light is directed to cavity 2, which is tuned and locked at resonance for another spatial mode, and so on.

In principle, it would be possible to sort out all modes (or groups of modes) having resonance lines within the resolution of the cavity. Therefore, it is possible to sort out a number of modes sufficient to build up quantum information schemes with dimensions greater than 2. 

The proposed scheme resembles the one implemented in Ref. \cite{Wei2020}. The main difference is the geometry of the cavity, which allows sorting out a higher number of HG modes. Also, the triangular design provides easy access to the reflected light without the need for optical isolation. This feature reduces losses along the cascade, improving fidelity (reducing cross-talk) in applications involving quantum (classical) information processing. This design can be complementary to that of Ref. \cite{Wei2020}, more suitable for sorting LG modes.

\section{Conclusion}
\label{sec:conc}

In conclusion, we demonstrate the use of an optical triangular cavity for discriminating between spatial paraxial modes. We illustrate the cavity performance presenting an experiment that uses a laser beam and a spatial light modulator to prepare superpositions of spatial modes and demonstrate the dynamical mode discrimination up to the fourth order in the Hermite-Gaussian modes basis. We also describe a proposal for the realization of a spatial mode sorter based on triangular cavities. 

The scheme is simple and reliable and can be associated to other available mode sorting techniques in Quantum Information applications.

\begin{acknowledgments}
The authors thank Giovani Pollachini for his technical contributions on the design of the cavity on the earlier stages of the experiment. We also acknowledge financial support from the Brazilian funding agencies Coordena\c{c}\~ao de Aperfei\c{c}oamento de Pessoal de N\'ivel Superior (CAPES), Conselho Nacional de Desenvolvimento Cient\'ifico e Tecnol\'ogico (CNPq) and Instituto Nacional de Ci\^encia e Tecnologia de Informa\c{c}\~ao Qu\^antica (INCT-IQ). 
\end{acknowledgments}

\bibliographystyle{apsrev}
\bibliography{SLM_and_cavity}

\end{document}